\documentclass[prr,twocolumn,floatfix,a4paper,superscriptaddress]{revtex4}

\usepackage{bm,color,amsmath,txfonts}
\usepackage{graphicx}
\usepackage{siunitx}
\usepackage{subfigure}
\usepackage{verbatim}
\usepackage{dcolumn}
\usepackage{bm}
\usepackage{epsf}
\usepackage{xcolor}
\usepackage{hyperref}
\usepackage{hhline}
\usepackage{float}
\usepackage{enumerate}
\usepackage{bbm}
\usepackage{lipsum}
\usepackage{mathrsfs}

\begin{document}
	
	\title{Optomagnonic generation of entangled travelling fields with different polarizations}
	
	\author{Zi-Xu Lu}
	\author{Huai-Bing Zhu}
	\author{Xuan Zuo}
	\author{Jie Li}\thanks{jieli007@zju.edu.cn}
	\affiliation{Zhejiang Key Laboratory of Micro-Nano Quantum Chips and Quantum Control \\ and School of Physics, Zhejiang University, Hangzhou 310027, China}

	\begin{abstract}
		The optomagnonic coupling between magnons and optical photons is an essential component for building remote quantum networks based on magnonics. Here we show that such a coupling, manifested as the magnon-induced Brillouin light scattering, can be exploited to entangle two propagating optical fields. The protocol employs two pairs of the whispering gallery modes coupled to the same magnon mode in a YIG microsphere or microdisk. In each pair a strong pump field is applied to activate either Stokes or anti-Stokes scattering. Due to the magnon mode involving in the two scattering processes and as a mediation, Stokes and anti-Stokes photons of different polarizations get entangled. The entanglement can be extracted by filtering the travelling output fields centered at the Stokes and anti-Stokes sidebands. Optimal conditions are identified under which strong output entanglement can be achieved.
	\end{abstract}
	
	\maketitle
	
	\section{Introduction}\label{I}
	
	Optical entanglement is a valuable quantum resource that enables a wide range of applications in quantum information science and technology, including quantum teleportation~\cite{Ze97,Furu98}, quantum network~\cite{K08,RH18}, quantum metrology~\cite{VG11}, quantum cryptography~\cite{PG02}, quantum logic operations~\cite{GJ01}, as well as in the fundamental tests of quantum mechanics~\cite{RH15,PG15,Ze15}.    To date,  multiple methods have proven effective in generating entangled optical fields, e.g., using the nonlinear parametric down-conversion (PDC) process~\cite{Ou92,YS95} and linear optical operations such as beam splitting~\cite{MA97,PK02}. The optical entanglement has been produced in many different systems, including nonlinear crystals~\cite{Ou92,YS95}, periodically poled lithium niobate waveguides~\cite{NG01}, quantum dots~\cite{YY00,AJ06}, atomic vapors~\cite{PD09}, coherent free electrons~\cite{IK22}, optomechanical systems~\cite{JM19}, and electro-optical devices~\cite{Qiu23}.  

	Here, we propose a new approach based on the recently developed optomagnonic system~\cite{Nakamura16,Zhang16,Haigh16,Nakamura18}. The system consists of an yttrium-iron-garnet (YIG) microsphere, or a YIG microdisk, which supports both optical whispering gallery modes (WGMs) and a magnetostatic magnon mode, and the optomagnonic interaction is manifested as the magnon-induced Brillouin light scattering (BLS). Such a system promises diverse potential applications in quantum information science and technology, such as quantum network~\cite{Jie21X,Zhou23,Tan25}, quantum teleportation~\cite{Fan23,Lu2}, quantum illumination~\cite{Zhou21}, Bell tests~\cite{Xie22}, the preparation of magnonic non-Gaussian~\cite{Bittencourt19,Qiong21,Jiang21,Lu1} and entangled~\cite{Zhu24,Song24} states, as well as in the microwave-to-optics conversion~\cite{NakamuraPRB,Op20,DongR,WWJ,Bei,GSA}.	 
	In contrast with most of the above theoretical proposals, which adopted a pair of WGMs and a single pump field, here our protocol involves two pairs of WGMs and two pump fields. In each pair, the two WGMs are of different polarizations, i.e., the transverse-magnetic (TM) and transverse-electric (TE)-polarized, and their frequency difference matches the magnon frequency, such that the optomagnonic scattering is resonantly enhanced. In one pair, a strong laser field is used to drive the TM-polarized WGM via e.g., a nanofiber coupled to the WGM resonator, to activate the optomagnonic Stokes scattering, while in the other, the TE-polarized WGM is pumped and the anti-Stokes scattering is activated. In the former, the Stokes scattering creates entanglement between the magnon mode and the TE-polarized Stokes photons, whereas in the latter, the anti-Stokes scattering realizes the state-swap interaction between the magnon mode and the TM-polarized anti-Stokes photons.  Given the fact that the magnon mode simultaneously involves the above two scattering processes, the Stokes and anti-Stokes photons get entangled due to the mediation of the common magnon mode. Since the generated Stokes and anti-Stokes photons are then coupled to and travelling in the fibers and mixed with the pump photons, two filters are used to extract the entanglement shared between the Stokes and anti-Stokes photons. We analyze optimal conditions of the system and filters for achieving strong entanglement of two output fields, and present numerical results which confirm our above analysis on the entanglement mechanism. 
	
	The paper is organized as follows. In Sec.~\ref{II}, we introduce the system and the protocol used to generate the entangled travelling optical fields. In Sec.~\ref{III}, we analyze the mechanism of entanglement generation, present the results of the optical entanglement, and show the impact of the couplings, dissipations, filter bandwidth, and bath temperature on the entanglement. Finally, we summarize the work in Sec.~\ref{IV}.

	\section{System and protocol}\label{II}
	
\begin{figure}[t]
		\includegraphics[width=0.95\linewidth]{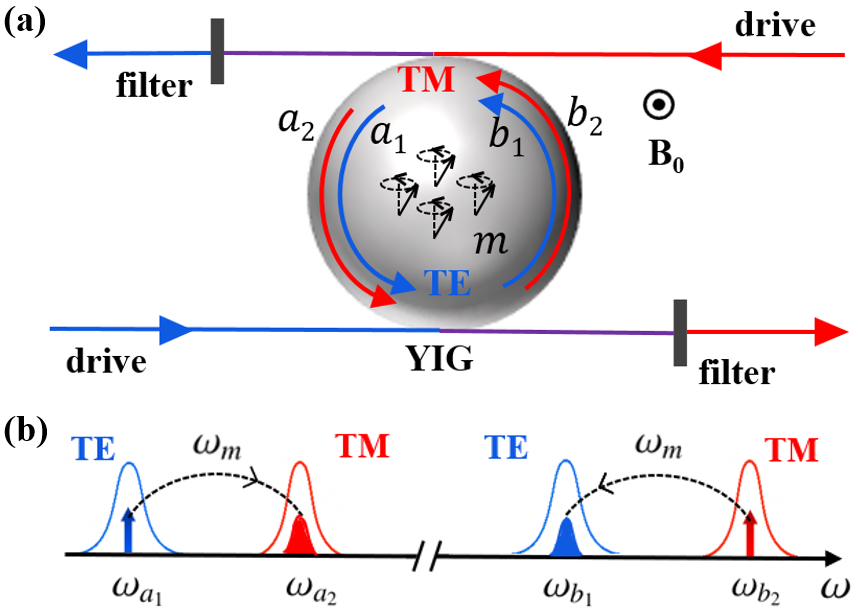}
		\caption{(a) Schematic of the protocol for generating entangled travelling fields. A YIG micro sphere or disk, placed in a uniform bias magnetic field, supports a magnetostatic magnon mode ($m$) and two pairs of WGMs ($a_{1,2}$ and $b_{1,2}$). In each pair, the optomagnonic interaction is manifested as the magnon-induced BLS. Two strong laser fields are used to drive the TM- and TE-polarized WGMs $b_{2}$ and $a_{1}$, via nanofibers coupled to the resonator, to simultaneously activate the Stokes and anti-Stokes scatterings. The generated Stokes and anti-Stokes photons are then coupled to and propagating in the fibers, and subsequently extracted from pump photons by two filters.  (b) Frequencies relation of the magnon mode, two pairs of WGMs, and two pump fields. The TM-polarized WGM $b_{2}$ is resonantly pumped to activate the Stokes scattering, creating entangled magnons and TE-polarized Stokes photons entering the WGM $b_{1}$. Similarly, the TE-polarized WGM $a_{1}$ is resonantly driven to activate the anti-Stokes scattering, producing TM-polarized anti-Stokes photons (via annihilating magnons) entering the WGM $a_{2}$.  The two vertical lines with arrows denote two pump fields. }
		\label{fig1}
	\end{figure}

	The optomagnonic system is depicted in Fig.~\ref{fig1}(a), which can be a YIG microsphere or microdisk.
	The single-photon optomagnonic coupling rate is typically weak, but the coupling strength can be greatly enhanced by a strong pump field. Due to the magnon-induced BLS, the photons in one WGM are scattered by lower-frequency magnons (typically in gigahertz), yielding two optical sidebands whose frequencies with respect to the WGM equal to the magnon frequency. The scattering probability is maximized when the scattered photons enter another WGM of the YIG micro sphere/disk, satisfying the so-called triple-resonance condition. Because of the selection rule~\cite{Sharma17,PAP17,Nakamuranjp,Haigh18} imposed by the conservation of the angular momenta of WGM photons and magnons, the BLS exhibits a pronounced asymmetry in the Stokes and anti-Stokes scattering strength. Besides, the selection rule causes different polarizations of the two WGMs, i.e., the TM- and TE-polarized WGMs~\cite{Nakamura16,Zhang16,Haigh16,Nakamura18}. We assume that the magnon-induced BLS occurs only between the TM and TE modes of the same WGM index, i.e., the orbital angular momentum of the WGM photons is conserved. In this case, the frequency of the TM mode is higher than that of the TE mode, $\omega_{\rm TM} > \omega_{\rm TE}$, due to the geometrical birefringence~\cite{Nakamura16,Nakamuranjp}. In the following, the two TE and TM modes of the same index are termed as a pair of WGMs.

	Our protocol for optical entanglement adopts two pairs of WGMs and a single magnon mode (Fig.~\ref{fig1}(a)). The frequency separation of the two pairs is tunable and depends on the frequencies of the optical fields we want to entangle. The Hamiltonian of the whole system is written as
	\begin{equation}
		H=H_{F}+H_{I}+H_{D},
	\end{equation}
	where $H_{F}/\hbar=\sum\omega_{O}O^{\dagger}O$ ($O=m,a_{1},a_{2},b_{1},b_{2}$) is the free Hamiltonian of the magnon mode and two pairs of WGMs, with $O$ ($O^{\dagger}$) being the annihilation (creation) operator and $\omega_{O}$ being the corresponding resonance frequency. Here, in each pair of WGMs, the triple-resonance condition is satisfied to maximize the scattering, i.e., $|\omega_{a_{2}}-\omega_{a_{1}}| \simeq |\omega_{b_{2}}-\omega_{b_{1}}| \simeq \omega_{m}$ (Fig.~\ref{fig1}(b)).  Without loss of generality, we assume that the subscript 2 (1) corresponds to the TM (TE)-polarized WGM. It is worth noting that although the geometry and material birefringence of a given YIG micro sphere/disk are fixed, the resulting frequency splitting between the TE and TM modes is not unique, i.e., it varies across different sets of mode indices~\cite{Nakamura16,Zhang16,Haigh16,Nakamura18}. Consequently, the optical spectrum inherently contains numerous TE-TM mode pairs, each with its own characteristic frequency difference. It is therefore practical to identify two mode pairs, whose intrinsic frequency differences naturally coincide or nearly coincide. We further note that the above triple-resonance condition does not have to be {\it exactly} satisfied. A small mismatch only reduces the strength of the Brillouin scattering, which, however, can be compensated by increasing the drive power.
	The Hamiltonian $H_{I}$ represents the optomagnonic interaction, corresponding to a three-wave process in each pair, given by~\cite{Jie21X,Bauer}
	\begin{equation}
		H_{I}/\hbar=g_{a} \left(a_{1}^{\dagger} a_{2} m^{\dagger} +a_{1} a_{2}^{\dagger} m \right)+ g_{b} \left(b_{1}^{\dagger} b_{2} m^{\dagger} +b_{1} b_{2}^{\dagger}m \right),
	\end{equation}
	where $g_{a(b)}$ are the bare optomagnonic coupling rates. To enhance the optomagnonic scattering, two strong laser fields are used to drive the TM mode $b_2$ and the TE mode $a_1$, respectively, and the corresponding driving Hamiltonian $H_{D}$ reads
	\begin{equation}
		H_{D}/\hbar=i\left(E_{1}a_{1}^{\dagger}e^{-i\omega_{p_{1}}t}+E_{2}b_{2}^{\dagger}e^{-i\omega_{p_{2}}t}- {\rm H.c.} \right),
	\end{equation}
	where $E_{1(2)}=\sqrt{2P_{1(2)}\kappa_{e}/\hbar\omega_{p_{1(2)}}}$ denotes the coupling strength between the WGM $a_{1}(b_{2})$ and the pump field with frequency $\omega_{p_{1(2)}}$ and power $P_{1(2)}$, with $\kappa_{e}$ being the external decay rate due to the coupling to one fiber which, for simplicity, is assumed identical for both the fibers.

	Consequently, the total Hamiltonian, with each pair of the WGMs in the frame rotating at its drive frequency, is given by
	\begin{equation}
		\begin{aligned}
			H/\hbar=&\omega_{m}m^{\dagger}m+\Delta_{a_{1}} a_{1}^{\dagger}a_{1}+\Delta_{a_{2}} a_{2}^{\dagger}a_{2}+\Delta_{b_{1}} b_{1}^{\dagger}b_{1}+\Delta_{b_{2}} b_{2}^{\dagger}b_{2}\\
			&+g_{a} \left(a_{1}^{\dagger} a_{2} m^{\dagger} +a_{1} a_{2}^{\dagger} m \right) + g_{b} \left(b_{1}^{\dagger} b_{2} m^{\dagger} +b_{1} b_{2}^{\dagger}m \right)\\
			&+i\left(E_{1}a_{1}^{\dagger}+E_{2}b_{2}^{\dagger}- {\rm H.c.}\right),
		\end{aligned}
	\end{equation}
	where $\Delta_{a_{1(2)}}=\omega_{a_{1(2)}}-\omega_{p_{1}}$ and $\Delta_{b_{1(2)}}=\omega_{b_{1(2)}}-\omega_{p_{2}}$. We consider the situation where the WGMs are resonantly pumped, i.e.,  $\Delta_{a_{1}}=\Delta_{b_{2}}=0$, and $\Delta_{a_{2}}=-\Delta_{b_{1}}=\omega_m$ (Fig.~\ref{fig1}(b)). This leads to the following simplified Hamiltonian
	\begin{equation}
		\begin{aligned}
			H/\hbar=&\omega_{m}\left(m^{\dagger}m+a_{2}^{\dagger}a_{2}-b_{1}^{\dagger}b_{1}\right)+g_{a} \left(a_{1}^{\dagger} a_{2} m^{\dagger} +a_{1} a_{2}^{\dagger} m \right)\\
			&+ g_{b} \left(b_{1}^{\dagger} b_{2} m^{\dagger} +b_{1} b_{2}^{\dagger}m \right)+i\left(E_{1}a_{1}^{\dagger}+E_{2}b_{2}^{\dagger}- {\rm H.c.}\right).
		\end{aligned}
	\end{equation}
	Since the WGMs $a_{1}$ and $b_{2}$ are strongly driven, which allows them to be treated classically as real numbers $\alpha_{1}\equiv\langle a_{1}\rangle \simeq E_{1}/(2\kappa_{e}+\kappa_{i})$ and $\beta_{2}\equiv\langle b_{2}\rangle \simeq E_{2}/(2\kappa_{e}+\kappa_{i})$~\cite{Jie21X} with $\kappa_{i}$ being the intrinsic decay rate of the WGM, which for simplicity is assumed identical for the WGMs involved. Thus, the total decay rate of each WGM includes two external decay rates due to the coupling to two fibers and the intrinsic decay rate, i.e.,  $\kappa_{\rm tot}= 2\kappa_{e}+\kappa_{i}$. The above Hamiltonian then becomes
	\begin{equation}
		\begin{aligned}
			H/\hbar=&\omega_{m}\left(m^{\dagger}m+a_{2}^{\dagger}a_{2}-b_{1}^{\dagger}b_{1}\right)+g_{a} \alpha_{1} \left( a_{2} m^{\dagger} + a_{2}^{\dagger} m \right)\\
			&+ g_{b} \beta_{2} \left(b_{1}^{\dagger} m^{\dagger} +b_{1} m \right),
		\end{aligned}
	\end{equation}
	which leads to the following effective Hamiltonian in the interaction picture
	\begin{equation}\label{eq:1}
		H_{\rm eff}/\hbar=G_{a}\left(a_{2}m^{\dagger}+a_{2}^{\dagger}m \right)+G_{b} \left(b_{1}^{\dagger}m^{\dagger}+b_{1}m \right),
	\end{equation}
	with $G_{a}=g_{a}\alpha_{1}$ and $G_{b}=g_{b}\beta_{2}$ being the pump-enhanced optomagnonic coupling strength, responsible for the $a_{2}$-$m$ state-swap interaction and the $b_{1}$-$m$ two-mode squeezing (TMS) interaction, respectively. 
	The corresponding quantum Langevin equations (QLEs), by including the dissipation and input noise of each mode, are given by
	\begin{equation}
		\begin{aligned}
			&\dot{m}=-\kappa_{m}m-iG_{a}a_{2}-iG_{b}b_{1}^{\dagger}+\sqrt{2\kappa_{m}}m^{\rm in},\\
			&\dot{a}_{2}=-\kappa_{\rm tot} a_{2}-iG_{a}m+ \sqrt{2\kappa_{e}}a_{2e_{1}}^{\rm in}+\sqrt{2\kappa_{e}}a_{2e_{2}}^{\rm in}+\sqrt{2\kappa_{i}}a_{2i}^{\rm in}, \\
			&\dot{b}_{1}=- \kappa_{\rm tot} b_{1}-iG_{b}m^{\dagger}+\sqrt{2\kappa_{e}}b_{1e_{1}}^{\rm in}+\sqrt{2\kappa_{e}}b_{1e_{2}}^{\rm in}+\sqrt{2\kappa_{i}}b_{1i}^{\rm in},  \\
		\end{aligned}
	\end{equation}
	where $\kappa_{m}$ ($m^{\rm in}$) is the decay rate (input noise) of the magnon mode, and $a_{2k}^{\rm in}$ ($k=e_{1},e_{2},i$) denote three independent input noises of the WGM $a_{2}$ due to the three dissipation channels mentioned above.  Similarly, $b_{1k}^{\rm in}$ are the three noise operators associated with mode $b_{1}$.  Since the internal dissipation of photons is typically high in magnets~\cite{Nakamura16,Zhang16,Haigh16,Nakamura18}, we assume a much larger intrinsic decay rate $\kappa_{i} = 4\kappa_{e} \equiv 4\kappa$, i.e., the total decay rate $\kappa_{\rm tot}=6\kappa$.  The input noises are zero-mean and the nonzero correlation functions are: $\langle m^{\rm in}(t)m^{\rm in\dagger}(t')\rangle=[N_{m}(\omega_{m})+1]\delta(t-t')$, $\langle m^{\rm in\dagger}(t)m^{\rm in}(t')\rangle=N_{m}(\omega_{m})\delta(t-t')$, $\langle a_{2k}^{\rm in}(t)a_{2k}^{\rm in\dagger}(t')\rangle=[N_{a_{2}}(\omega_{a_{2}})+1]\delta(t-t')$, $\langle a_{2k}^{\rm in\dagger}(t)a_{2k}^{\rm in}(t')\rangle=N_{a_{2}}(\omega_{a_{2}})\delta(t-t')$, and $\langle b_{1k}^{\rm in}(t)b_{1k}^{\rm in\dagger}(t')\rangle=[N_{b_{1}}(\omega_{b_{1}})+1]\delta(t-t')$, $\langle b_{1k}^{\rm in\dagger}(t)b_{1k}^{\rm in}(t')\rangle=N_{b_{1}}(\omega_{b_{1}})\delta(t-t')$~\cite{Zoller}, with $N_{j}(\omega_{j})$=[exp$(\hbar\omega_{j}/k_{B}T)-1]^{-1}$ ($j=m,a_{2},b_{1}$) being the equilibrium mean thermal excitation number of the corresponding mode, where $k_{B}$ is the Boltzmann constant and $T$ is the bath temperature.

	We note that the effective Hamiltonian~\eqref{eq:1} can already indicate an effective TMS interaction between the WGMs $a_{2}$ and $b_{1}$, which gives rise to entanglement of two intracavity fields. However, intracavity entanglement is inaccessible, and what can be accessed is that of the output fields propagating in the fibers.   
	The output field of the WGM $a_{2}$ ($b_{1}$) in one specific fiber (Fig.~\ref{fig1}(a)) can be obtained via the input-output relation: 
	\begin{equation}
		\begin{aligned}
			a_{2}^{\rm out}(t)&={\sqrt{2\kappa_{e}}} a_{2}(t)-a_{2e_{2}}^{\rm in}(t), \\ 
			b_{1}^{\rm out}(t)&={\sqrt{2\kappa_{e}}} b_{1}(t)-b_{1e_{1}}^{\rm in}(t), 
		\end{aligned}
	\end{equation}
	which form {\it continuous} output spectra in the frequency domain. Therefore, filters are typically required to define two output {\it modes} and extract the quantum correlation shared between the Stokes and anti-Stokes photons in the two propagating fields. The output modes are defined in the following way~\cite{DV08}
	\begin{equation}
		\begin{aligned}
			&A_{2}^{\rm out}(t)=\int_{-\infty}^{t}ds F_{2}(t-s)a_{2}^{\rm out}(s),\\
			&B_{1}^{\rm out}(t)=\int_{-\infty}^{t}ds F_{1}(t-s)b_{1}^{\rm out}(s),
		\end{aligned}
	\end{equation}
	where $F_{j}(t)$ ($j=1,2$) is the filter function, which, for the simplest square filter, is given by 
	\begin{equation}
		F_{j}(t)=\frac{h(t)-h(t-\tau_{j})}{\sqrt{\tau_{j}}}e^{-i\Omega_{j}t}.
	\end{equation}
	Here, $h(t)$ denotes the Heaviside step function, and $\Omega_{j}$ ($\tau_{j}$ or $1/\tau_{j}$) is the central frequency (time duration or frequency bandwidth) of the $j$th filter. We note that in each optical path of the fiber a filter is used to select either Stokes or anti-Stokes photons with different frequencies and polarizations.

	Due to the Gaussian nature of the noises and the linearized dynamics, the state of the magnon mode and two output modes is a three-mode Gaussian state~\cite{JieL,DV07}, which is completely characterized by a $6\times6$ covariance matrix (CM) $V^{\rm out}$ with its entries defined as
	\begin{equation}
		V_{ik}^{\rm out}(t)=\dfrac{1}{2}\left\langle u_{i}^{\rm out}(t)u_{k}^{\rm out}(t)+u_{k}^{\rm out}(t)u_{i}^{\rm out}(t)\right\rangle,
	\end{equation}
	where $u^{\rm out}(t)=[X_{m}(t),Y_{m}(t),X_{A_{2}}^{\rm out}(t),Y_{A_{2}}^{\rm out}(t),X_{B_{1}}^{\rm out}(t),Y_{B_{1}}^{\rm out}(t)]^{T}$ is the vector of the quadratures of the magnon and two output modes which are defined as $X_{m}(t)=[m(t)+m(t)^{\dagger}]/\sqrt{2}$, $Y_{m}(t)=i[m(t)^{\dagger}-m(t)]/\sqrt{2}$, and $X_{l}^{\rm out}(t)=[l^{\rm out}(t)+l^{\rm out}(t)^{\dagger}]/\sqrt{2}$, $Y_{l}^{\rm out}(t)=i[l^{\rm out}(t)^{\dagger}-l^{\rm out}(t)]/\sqrt{2}$ ($l=A_{2},B_{1}$). 
	Under appropriate conditions, the system reaches a steady state, resulting in stationary output entanglement. In this situation, the stationary CM of the system can be conveniently achieved in the frequency domain by taking Fourier transforms~\cite{DV08}.  After some calculations, we obtain the stationary CM in the form
	\begin{equation}
		V^{\rm out}=\int d\omega \tilde{T}(\omega)\left(\tilde{M}(\omega)+P^{\rm out} \right)D(\omega)\left(\tilde{M}(\omega)^{\dagger}+P^{\rm out} \right)\tilde{T}(\omega)^{\dagger},
	\end{equation}
	where $P^{\rm out}={\rm Diag}\left[0, 0, (2\kappa)^{-1},(2\kappa)^{-1},(2\kappa)^{-1},(2\kappa)^{-1} \right]$ is the projector onto the four-dimensional space associated with the output quadratures, the diffusion matrix $D(\omega)={\rm Diag} \Big[\kappa_{m}\left(2N_{m}{+}\,1\right),\kappa_{m}\left(2N_{m}{+}\,1\right),4\kappa\left(2N_{a_{2}}{+}\,1\right), 4\kappa\left(2N_{a_{2}}{+}\,1\right),$ $4\kappa\left(2N_{b_{1}}+1\right),4\kappa\left(2N_{b_{1}}+1\right) \Big]$, $\tilde{T}(\omega)$ is in the form 
	\begin{widetext}
		\begin{equation}
			\begin{aligned}
				\tilde{T}(\omega)=\begin{pmatrix}\dfrac{1}{\sqrt{2\pi}}&0&0&0&0&0\\0&\dfrac{1}{\sqrt{2\pi}}&0&0&0&0\\0&0&\sqrt{2\kappa}\:\mathrm{Re}\:F_{2}(\omega)&-\sqrt{2\kappa}\:\mathrm{Im}\:F_{2}(\omega)&0&0\\0&0&\sqrt{2\kappa}\:\mathrm{Im}\:F_{2}(\omega)&\sqrt{2\kappa}\:\mathrm{Re}\:F_{2}(\omega)&0&0\\0&0&0&0&\sqrt{2\kappa}\:\mathrm{Re}\:F_{1}(\omega)&-\sqrt{2\kappa}\:\mathrm{Im}\:F_{1}(\omega)\\0&0&0&0&\sqrt{2\kappa}\:\mathrm{Im}\:F_{1}(\omega)&\sqrt{2\kappa}\:\mathrm{Re}\:F_{1}(\omega)\end{pmatrix},
			\end{aligned}
		\end{equation}
	\end{widetext}
	with $F_{1(2)}(\omega)$ being the Fourier transform of filter function $F_{1(2)}(t)$, and 
	\begin{equation}
		\tilde{M}(\omega)=(i\omega+A)^{-1},
	\end{equation}
	with the drift matrix
	\begin{equation}
		\begin{aligned}
			A=\begin{pmatrix}-\kappa_{m}&0&0&G_{a}&0&-G_{b}\\0&-\kappa_{m}&-G_{a}&0&-G_{b}&0\\0&G_{a}&-\kappa_{\rm tot}&0&0&0\\-G_{a}&0&0&-\kappa_{\rm tot}&0&0\\0&-G_{b}&0&0&-\kappa_{\rm tot}&0\\-G_{b}&0&0&0&0&-\kappa_{\rm tot}\end{pmatrix}.
		\end{aligned}
	\end{equation}
	The stability of the system is guaranteed by the negative eigenvalues (real parts) of the drift matrix $A$, and all the results presented in this work are in the steady state.
	
	Once the stationary $V^{\rm out}$ is achieved, we extract the $4\times4$ CM $V_{4}$ associated with the two output modes, which takes the form of $V_{4}=[V_{A},V_{AB};V_{AB}^{T},V_{B}]$, with $V_{A}$, $V_{B}$ and $V_{AB}$ being $2\times2$ blocks.  We adopt  the logarithmic negativity~\cite{Vidal} $E_{N}$ to quantify the entanglement between two output modes, which is defined as
	\begin{equation}
		E_{N}\equiv\max\left[0,-\ln 2\eta^{-}\right],
	\end{equation}
	where $\eta^{-}\equiv2^{-1/2}\{\Sigma(V_{4})-[\Sigma(V_{4})^{2}-4\det V_{4}]^{1/2}\}^{1/2}$, with $\Sigma(V_{4})\equiv\det V_{A}+\det V_{B}-2\det V_{AB}$.

	\begin{figure}[b]
		\includegraphics[width=\linewidth]{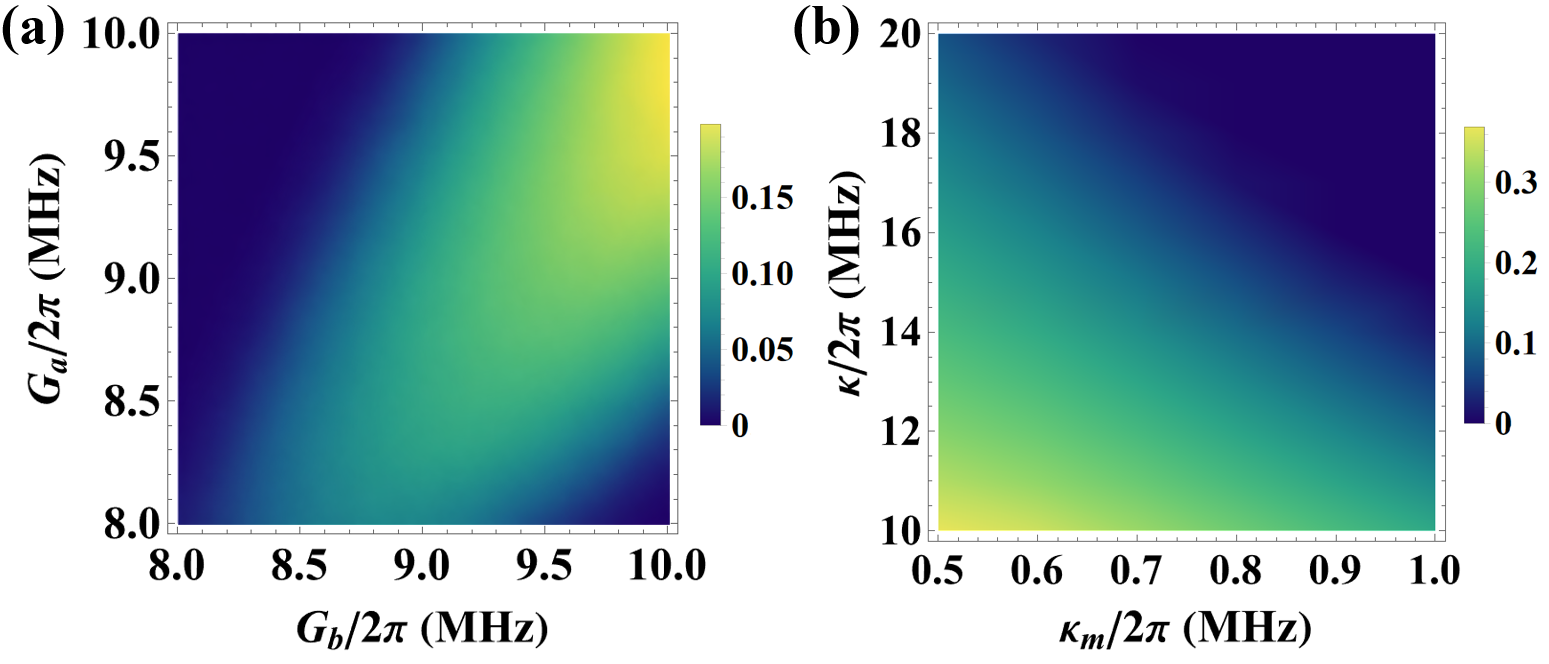}
		\caption{Stationary entanglement $E_{N}$ of two output modes versus (a) effective optomagnonic coupling rates $G_{a}$ and $G_{b}$; and (b) decay rates $\kappa_{m}$ and $\kappa$. In (b), we fix {$G_{a}/2\pi=G_{b}/2\pi=10$ MHz}. The other parameters are provided in the text.}
		\label{fig2}
	\end{figure}
	
	\section{Results}\label{III}

	The determination of two optimal {optomagnonic} couplings $G_{a}$ and $G_{b}$ is important for achieving strong output entanglement. As revealed from the Hamiltonian~\eqref{eq:1}, the output entanglement is a result of the simultaneous activation of both Stokes and anti-Stokes scatterings, in which the former gives rise to the TMS interaction between modes $b_{1}$ and $m$ with strength $G_{b}$, while the latter yields the state-swap interaction between modes $a_{2}$ and $m$ with strength $G_{a}$. The trade-off between these two interactions yields an optimal ratio of the two couplings. In the limit case of $G_{b}\gg G_{a}$, i.e., the TMS interaction is much stronger than the state-swap interaction, the former would generate strong entanglement between $b_{1}$ and $m$ (assume the system remains stable), however, due to the slow state swapping rate between $a_{2}$ and $m$, the created entanglement cannot be efficiently transferred to mode $a_{2}$, leading to weak output entanglement. In the opposite case of $G_{a}\gg G_{b}$, although the state swapping rate between $a_{2}$ and $m$ is fast, a weak TMS interaction creates weak entanglement between $m$ and $b_{1}$, which also corresponds to weak output entanglement.  Optimal coupling rates should be away from these two extreme cases. This is confirmed by Fig.~\ref{fig2}(a), which indicates that the optimal situation corresponds to 
	two close couplings 	$G_{a} \simeq G_{b}$.
	In getting Fig.~\ref{fig2}(a), we have used the following feasible parameters: $\omega_{m}/2\pi=10$~GHz, $\omega_{a_{2}}/2\pi=193404.5$ GHz, $\omega_{b_{1}}/2\pi=193174.5$ GHz, $\kappa_{m}/2\pi=0.5$ MHz, $\kappa/2\pi=15$~MHz, $g_{a(b)}/2\pi=1$ kHz, and the system is placed at a low bath temperature $T=0.01$~K. The filter parameters are chosen as $\Omega_{1(2)}=\omega_{b_{1}(a_{2})}$ and $\tau_{1(2)}\equiv\tau=10$ $\mu s$.  Note that for macroscopic submillimeter YIG spheres used in Ref.~\cite{Nakamura16,Zhang16,Haigh16}, the bare coupling $g$ is very weak, $\sim10$ Hz, which requires a high drive power (tens of mW) to get a MHz effective coupling $G$, which will cause considerable heating effects.   A more feasible system would be a YIG microdisk, in which the bare coupling can be significantly improved by reducing the mode volume and increasing the mode overlap. For example, a theory predicts that $g$ can reach 1 kHz for a YIG disk with radius of 20~$\mu$m~\cite{Belotelov}. This will greatly reduce the power to the order of ~10 $\mu$W to have a MHz level coupling $G$.

	Figure~\ref{fig2}(b) studies the influence of the optical and magnon decay rates on the entanglement. Clearly, the entanglement reduces with both the decay rates.  Since the magnon dissipation rate is intrinsic and dominated by two-magnon scattering from lattice and surface defects~\cite{Chumak1,Chumak}, there is not much room for improvement. A recent experiment reported one of the lowest magnon decay rates, $\kappa_{m}/2\pi = 0.5$ MHz~\cite{Shen25}. Therefore, reducing the optical dissipation of the WGMs, especially the significant intrinsic loss, would be more practical to improve the entanglement.

	
	\begin{figure}[t]
		\includegraphics[width=\linewidth]{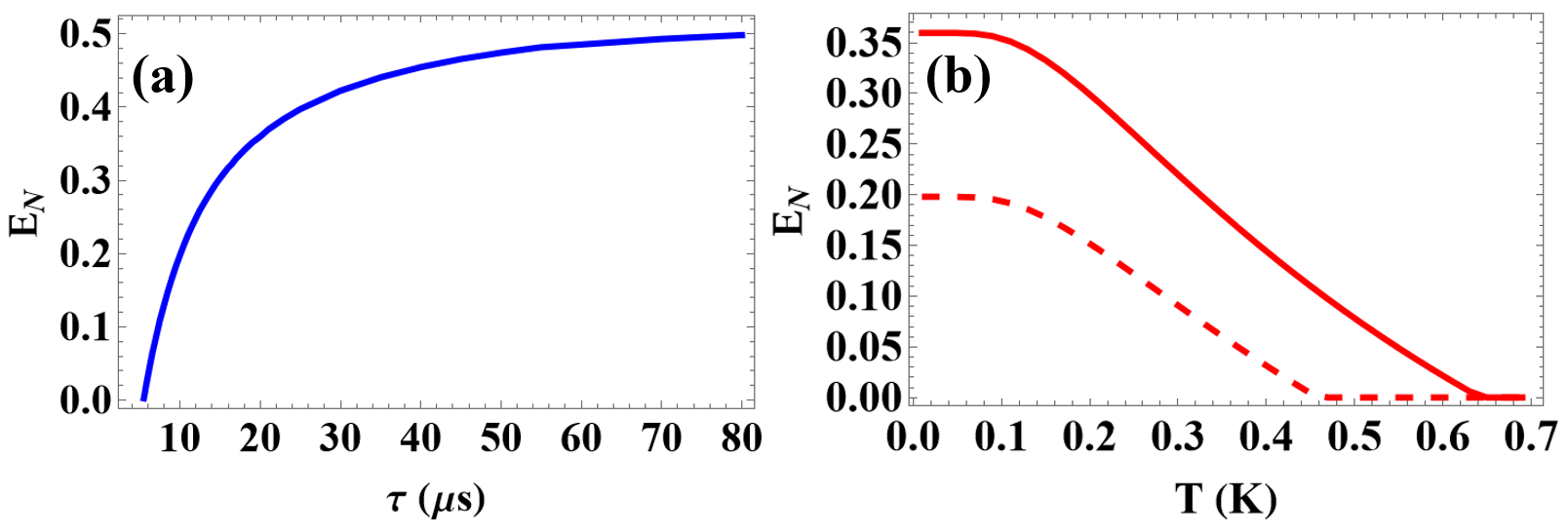}
		\caption{Stationary output entanglement $E_{N}$ versus (a) filter time duration $\tau$; and (b) {bath temperature $T$. In both plots, we take $G_{a}/2\pi=G_{b}/2\pi=10$ MHz, and the solid (dashed) line in (b) is for $\tau=20$ $\mu$s (10 $\mu$s).}  The other parameters are the same as in Fig.~\ref{fig2}(a).}
		\label{fig3}
	\end{figure}

	The effect of the filter parameters on the entanglement is straightforward. {Since the output entanglement originates from the generated Stokes and anti-Stokes photons, the central frequencies of the filters being resonant with the Stokes and anti-Stokes sidebands, i.e., $\Omega_{1(2)}=\omega_{b_{1}(a_{2})}$ used in Fig.~\ref{fig2}, corresponds to the optimal situation of extracting the generated entanglement~\cite{DV08}.} 
	In Fig.~\ref{fig3}(a), we plot the entanglement $E_{N}$ as a function of the filter time duration $\tau$. Clearly, the longer the duration is, or the narrower the filter bandwidth is, the better. Due to the significant intrinsic loss, the time duration should be sufficiently long, $\tau >5.5$ $\mu$s, to obtain any entanglement in the output fields.
	{The entanglement exhibits a saturation effect when increasing the filter duration. This is due to the fact that the act of filtering introduces back-action that fundamentally limits the maximum attainable entanglement.}
	In Fig.~\ref{fig3}(b), we further show the stationary output entanglement $E_{N}$ versus the bath temperature $T$ for two duration times $\tau=10$ $\mu$s and $20$ $\mu$s. We see that the optical entanglement is robust against bath temperature with a nonzero $E_N$  being up to 0.65 K (0.46 K) for $\tau=20$ $\mu$s (10 $\mu$s). This is because on the one hand, the magnon frequency (in gigahertz) is relatively high, thus less affected by thermal noise; on the other hand, our protocol contains a cooling mechanism~\cite{TJK07,FM07}, i.e., the anti-Stokes scattering.

	\section{Conclusions and Discussions}\label{IV}

	In conclusion, we have presented a promising efficient approach for generating stationary entanglement between two travelling optical fields based on the magnon-induced BLS. By adopting two pairs of the WGMs coupled to a single magnon mode and simultaneously activating the optomagnonic Stokes and anti-Stokes scatterings, the generated Stokes and anti-Stokes photons get entangled due to the mediation of the magnon mode, which can be directly accessed by filtering the two propagating output fields. The degree of entanglement can be enhanced by reducing the filter bandwidth and further improved by distilling the entanglement using, e.g., non-Gaussian operations~\cite{Lu2}. {The proposed entanglement generator based on a YIG microdisk coupled to fibers can be integrated with the fiber-based magnonic quantum network~\cite{Jie21X}, finding applications in quantum state transfer and entanglement distribution among remote quantum nodes. The optical entanglement also finds a wide range of applications in quantum information science and quantum sensing, including continuous-variable quantum teleportation~\cite{Furu98}, quantum illumination~\cite{ill}, and improving the detection sensitivity~\cite{Xia}.}
	
	It is worth noting that our approach exhibits advantages over existing methods~\cite{Ou92,YS95,MA97,PK02,NG01,YY00,AJ06,PD09,IK22,JM19,Qiu23}. First, the frequencies of the entangled optical fields are tunable due to two unique features of the proposed optomagnonic system: The frequency separation of two pairs of WGMs is tunable, and the magnon frequency is tunable and proportional to the bias magnetic field. Second, it provides an inherent microwave-optics interface. Since the entanglement is mediated by a magnon mode, which can couple to microwave photons, our protocol can be extended to generate microwave-optics entanglement by further coupling magnons to microwave fields.

	\section*{ACKNOWLEDGMENTS}
	This work was supported by Zhejiang Provincial Natural Science Foundation of China (Grant No. LR25A050001), National Natural Science Foundation of China (Grant No. 12474365, 92265202) and National Key Research and Development Program of China (Grant No. 2024YFA1408900, 2022YFA1405200).

\end{document}